# Deutsch Algorithm on Classical Circuits

Assist.Prof.Dr. Osman Kaan EROL

Istanbul Technical University, Electrical-Electronics Faculty, Computer Engineering Dept.

Istanbul-Turkey

**Abstract:**

The well-known Deutsch Algorithm (DA) and Deutsch-Jozsha Algorithm (DJA) both are used as an evidence to the power of quantum computers over classical computation mediums. In these theoretical experiments, it has been shown that a quantum computer can find the answer with certainty within a few steps although classical electronic systems must evaluate more iterations than quantum computer. In this paper, it is shown that a classical computation system formed by using ordinary electronic parts may perform the same task with equal performance than quantum computers. DA and DJA quantum circuits act like an analog computer, so it is unfair to compare the bit of classical digital computers with the qubit of quantum computers. An analog signal carrying wire will of course carry more information that a bit carrying wire without serial communication protocols.

**I Deutsch Algorithm**

In Nielsen's and Chuang's book [1], Deutsch Algorithm (DA) is considered as a challenge over classical computers. DA poses a challenge to determine with certainty the character of a logic function $f$ having one or more inputs and one output logic variable. For DA this function is assumed to give either a constant value ($f(0) = f(1) = 0$ or $1$) or balanced output ($f(0) \neq f(1)$). The challenging part of this experiment is said to lie in using only one evaluation of $f$. In classical computing theory, it is required to perform at least two measurements for this experiment to precisely make the distinction between constant or balanced type function is used. The possible cases for the function $f$ are summarized in table 1.1

Table 1.1 The three possible logic functions for Deutsch Algorithm

| Function | Description |
|---|---|
| 0 | Constant output with logic level 0 |
| 1 | Constant output with logic level 1 |
| Balanced | $f(0) \neq f(1)$ |

A constant output function $f$ is easy to realise: The output is tied to ground level or logic '1' level. The input signal is discarded. In this case, the output is independent to the input.

A balanced type function *f* is easy to realise too: The input to output relation can be either to forward the input signal to output or forward the complement of the input signal. All 4 types of the possible *f* functions are given in table 1.2.

Table 1.2 The four possible logic functions for Deutsch Algorithm

| Function | Description |
|---|---|
| 0 | Output = 0 |
| 1 | Output = 1 |
| Balanced – non inverting | *f*(0) = 0, *f*(1) = 1 |
| Balanced - inverting | *f*(0) = 1, *f*(1) = 0 |

Quantum circuit implementation of Deutsch Algorithm is given Figure 1.1.

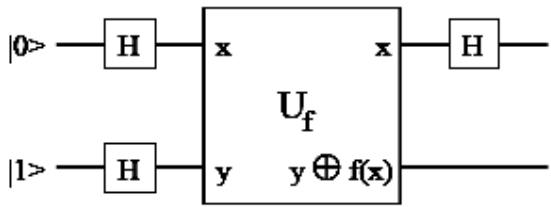

Figure 1.1 Quantum Circuit for Deutsch Algorithm

In this circuit, both inputs *x* and *y* to the block $U_f$ receive one qubit of data having a state neither pure state |0⟩ nor pure state |1⟩ but in between. This state is called "bell-state" and the characteristic of this state is that a qubit will generate 0 or 1 of equal probability after measurement. The final result measured at the output of the Hadamard gate at the output shows with certainty the nature of *f(x)*. A 0 at the output will prove *f(x)* is of constant type while a 1, will show *f(x)* is of balanced type.

By using classical digital computers, to determine the nature of *f*, one should make two measurements: First after applying logical 0 to the input of *f*, then secondly, after applying logical to the input of *f*. Finally it is easy to determine whether *f* is produced constant or balanced output. One single measurement and hence one single evaluation of *f* will not give enough information to determine whether *f* is of constant or balanced type.

But what about analogue computers?

**II Deutsch Algorithm (DA) on analogue classical computers.**

A state ψ of a qubit can be expressed as a superposition of base states [1]:

$$|\psi\rangle = \alpha |0\rangle + \beta |1\rangle \qquad 2.1$$

and

$$\alpha^2 + \beta^2 = 1 \qquad 2.2$$

A representation of a single qubit is Bloch Sphere [1]. Bloch Sphere is a unit sphere due to Eq 2.2. Here second norm of distance is used. Different types of unit spheres according the metrics used are given Fig 2.1.

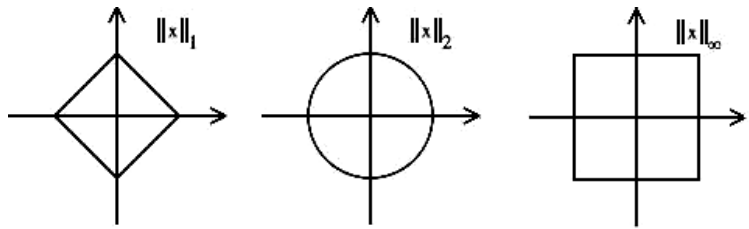

Fig 2.1 Some unit spheres

‖x‖$_1$ norm will be used throughout this study to express median logic levels. For example, an analogous bell state will be obtained by setting the line voltage to a level in between logic 0 and logic 1 voltages. If 5V logic is used, then a bell state will be obtained by using 2.5 V.

Classical logic gates have voltage transitions very fast. It is therefore difficult to obtain a stable median voltage level for all circuit and for every time of operation. Without giving sacrifice on the operation around logic levels, buffer and inverter gates will be obtained by using ordinary op-amps and by setting their gain to the unity. The black box *f* will be designed according to the Figure 2.2:

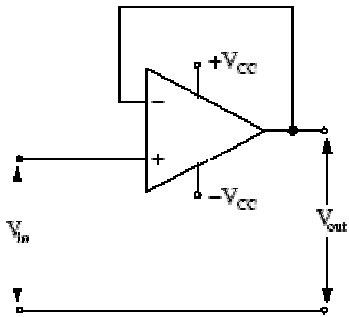

Figure 2.2 (a) A buffer circuit having transfer gain = 1

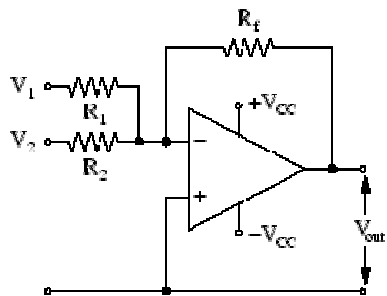

Figure 2.2 (b) An inverting circuit.

Normally it is also possible not to use Figure 2.2 (a) and tie the output to the input. In Figure 2.2 (b), inverting is possible if the input voltage is applied to any input of the adder and a constant voltage (For 5V logic, -5V, for 3.3V logic, -3.3 V) to the other input. Here all there resistors are set to be equal (For example 10K).

The circuits given in Figure 2.2 will perform as classical logic counterparts and generate logic signals conform to logic levels. The difference is that the circuit used in this study will accept intermediate voltage levels. Their gain is plus or minus unity. Therefore their output will swing freely in between logic 0 and logic 1 voltage levels. These transition voltages will be used to obtain a super-position of two logic states through only one wire and at the same time.

Our goal is to determine with certainty whether this function produce a constant output or balanced output by performing only one measurement.

If a constant output function has been chosen, then the output will be constant with voltage level conform to logic 0 or 1 (e.g. ground level or +5V). On the contrary, if a balanced function is chosen, then the output will swing from 0 to 1 or vice versa according to the logic input and function type. If a voltage that puts the input in a transition state is applied, (when it is driven by superposed data, or 2.5V in 5V logic), then at the output an intermediate voltage level will be obtained (e.g. 2.5V). It is therefore possible to determine the nature of this black-box (*f* function), by checking the voltage level at the output with only one measurement and only one function evaluation.

**III Deutsch - Jozsa Algorithm on analogue classical computers.**

Deutsch – Jozsa Algorithm (DJA) is a generalized version of the previous algorithm. In DJA, $f(\vec{x})$ is an *n* input one output logic function. It is assumed that *f* generates either constant output for all $\vec{x}$'s or balanced output, that is 0 for half of the possible inputs and 1 for the other half. Our goal is to find out which case *f* belongs to by performing only one measurement.

The quantum circuit for DJA is given below:

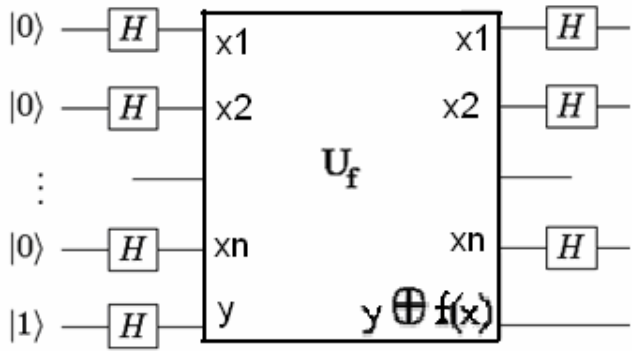

Figure 3.1 Quantum Circuit for Deutsch - Jozsa Algorithm

To implement the algorithm by using an analogous classical circuit, it is required to define logical AND and logical OR gates in a similar way given in Figure 2.2 (a) and (b).

Logic AND equivalent analog circuit can be defined as an analog multiplier circuit. The analog output is the multiplication of the analog inputs. Logic 1 is assumed to be given by means of a voltage level of 1V. There many IC manufacturers to offer many versions of analog multiplication circuits. Any one having a linear output response in the [0,1] V interval can be used for this purpose. It is important to note that the transfer gain of the multiplier must be or set to unity.

Logic OR equivalent analog circuit is an adder circuit + an optional limiter. The analog output is the sum of the analog inputs upper-limited to 1V if logic 1 is assumed to be given by means of a voltage level of 1V. Like multiplier circuits, analog adder circuit can be composed by means of op-amps or by using any adder IC. The output limitation can be done as given in [2]. The multiplier circuit will not need a limiter since the output can not be higher than 1V if the input voltages are within [0,1] V interval.

By using the inverter, logic equivalent AND and logic equivalent OR circuits, it is possible realize any logic function with $n$ inputs and one output.

To precisely determine the type of the function $f$, it is therefore required to set all the inputs to logic 0.5V. any logic function can be realised by the sum of product terms or first canonical representation such as [3-4]:

$$f = x_1 x_2 x_3 .. x_n + x_1' x_2 x_3' ... x_n + ....  \quad (3.1)$$

In the example given in Eq. 3.1, no simplification are carried out that is the product terms are composed of all input variables either by direct inclusion or by their logical complement. In this case, the output of the multiplier circuits taking part inside of $f$ will drop to a voltage level below 1V according to the number of inputs. However this is predefined: If there are $n$ inputs, the output of any product term will be either 0V or

$$V_{out} = \frac{1}{2^n} \text{ V} \quad (3.2)$$

The adder circuit must have $2^{n-1}$ inputs in order to generate a balanced output. Summing up all the outputs of the multiplier circuits will therefore yield to:

$$V_{out} = \frac{2^{n-1}}{2^n} = 0.5\text{V} \quad (3.3)$$

Any observer having a multi-meter will precisely decide whether $f$ has a constant output if he measures 0 or 1V, or $f$ is balanced if he measures 0.5V with <u>only one</u> measurement

**IV Conclusion**

A qubit is a data that can carry infinite number of information (state) in $O(n^3)$, here $n$ denotes the freedom axis in three dimensional space. A spin of a qubit can be positioned in any three angles, up or down (base states $|0\rangle$ and $|1\rangle$) being only two of them. In the presented study, a qubit, superposed data in other terms, can carry information in $O(n)$ since the distance metric used is the norm 1. Anyway, both approaches may have infinite number of states. Both signals can deliver logic 0 or logic 1 with a probability conform to their states. A qubit necessitates a quantum circuit or computer while an analogue electrical signal may be applied to classical analogue computer systems.

So for DA and DJA, a quantum computer plays a role like an analog computer, and an isomorphism can be shown between opamp and other analog circuits and quantum DA and DJA circuits.